\documentclass[10pt]{article}
\usepackage[a4paper,left=24mm,right=24mm,top=23mm,bottom=24mm]{geometry}
\usepackage[T1]{fontenc}
\usepackage{amsmath,amssymb,amsfonts,mathtools,amsthm}
\usepackage{txfonts}
\usepackage{booktabs,array,enumitem,microtype,mathrsfs}
\usepackage[numbers,sort&compress]{natbib}
\usepackage{needspace}
\usepackage{capt-of} 
\usepackage{url}
\usepackage[hidelinks,hypertexnames=false]{hyperref}
\hypersetup{
  bookmarksnumbered=true,
  pdfauthor={Yangcheng Li and Pingzhi Yuan},
  pdftitle={Self-Orthogonal MDS Principal-Ideal Codes},
  pdfsubject={self-orthogonal GRS generators and non-GRS MDS codes}
}
\numberwithin{equation}{section}
\newtheorem{theorem}{Theorem}[section]
\newtheorem{proposition}[theorem]{Proposition}
\newtheorem{lemma}[theorem]{Lemma}
\newtheorem{corollary}[theorem]{Corollary}

\newtheorem{remark}[theorem]{Remark}
\newtheorem{example}[theorem]{Example}

\newcommand{\F}{\mathbb F}
\newcommand{\PP}{\mathbb P}

\newcommand{\wt}{\operatorname{wt}}

\newcommand{\cC}{\mathcal C}

\begin{document}

\title{Self-Orthogonal MDS Principal-Ideal Codes}
\author{Yangcheng Li\thanks{Corresponding author. Email: \texttt{liyc@m.scnu.edu.cn}.}\quad
Pingzhi Yuan\thanks{Email: \texttt{yuanpz@scnu.edu.cn}.}\\[3pt]
\small School of Mathematical Sciences, South China Normal University\\
\small Guangzhou 510631, Guangdong, China}
\date{}
\maketitle

\begin{abstract}
	We investigate Euclidean self-orthogonality and the generalized
	Reed--Solomon (GRS) property for coefficient-vector codes generated
	by a monic polynomial $g$ and its shifts
	$xg,\ldots,x^{k-1}g$, of length $n=r+k$ and dimension $k$,
	where $\deg g=r$ and $3\le k\le r$.
	We give a complete classification, over any algebraically closed
	field, of the generators for which these codes are both
	self-orthogonal and GRS.
	Such generators have either distinct roots in a geometric
	progression satisfying explicit order and scalar conditions,
	or a single repeated root in certain positive characteristics.
	In particular, only finitely many generators occur for fixed
	$r$, $k$, and characteristic.
	We determine their maximal self-orthogonal lengths, show that
	the corresponding maximal-length codes are cyclic or negacyclic,
	and obtain exact generator counts over finite fields, together
	with the least fields of definition.
	In contrast, when $3\le k<r$, we prove that every algebraically
	closed field supports infinitely many pairwise monomially
	inequivalent self-orthogonal non-GRS maximum distance separable
	(MDS) codes in this family.
	Consequently, arbitrarily many such codes can be realized over
	a common finite extension of any prescribed finite field,
	while remaining inequivalent and non-GRS over its algebraic
	closure.
\end{abstract}



\section{Introduction}

Let $g=\sum_{j=0}^r g_jx^j$ be monic of degree $r\ge1$ over a field
$K$, and fix $k\ge1$. Put $n=r+k$ and let
$\operatorname{coeff}_{<n}(w)$ denote the coefficient vector of a
polynomial of degree less than $n$. We study
\begin{equation}\label{eq:code}
	\cC_g(n)
	=\{\operatorname{coeff}_{<n}(ug):u\in K[x],\ \deg u<k\}
	\subseteq K^n.
\end{equation}
The generating rows $g,xg,\ldots,x^{k-1}g$ have distinct degrees,
so the code has dimension $k$ and Singleton bound $d\le r+1$
\cite{Singleton}. We call $g$ an MDS generator when equality holds.
For any monic $h$ of degree $k$, \eqref{eq:code} is also the
coefficient representation of the principal ideal $(g)$ in
$K[x]/(gh)$ \cite[Proposition~7.1]{LiYuanOrbit}, explaining the
term \emph{principal-ideal code}.

A generalized Reed--Solomon (GRS) code is a weighted evaluation
code at distinct points of the projective line, allowing infinity.
Coordinate permutations and nonzero coordinate scalings define
\emph{monomial equivalence}, which preserves distance and GRS type.
Euclidean self-orthogonality requires
$\sum_i v_iw_i=0$ for every $v,w\in C$ and depends on the coordinate
scalings. We impose it in the coefficient coordinates of
\eqref{eq:code}. Our aim is to classify the generators giving
self-orthogonal GRS codes and to determine how many inequivalent
self-orthogonal MDS codes occur outside the GRS class.
We focus on $3\le k\le r$. Self-orthogonality forces $k\le r$,
while over an algebraically closed field every MDS code of
dimension one or two is GRS. Remark~\ref{rem:dimension-range}
explains these boundary cases over an arbitrary field.

The classical evaluation construction is due to Reed and Solomon
\cite{ReedSolomon}. Jin and Xing constructed MDS self-dual codes
from GRS codes \cite{JinXing}, and Fang, Liu and Luo characterized
the weights giving Euclidean self-orthogonality \cite{FangLiuLuo}.
Related work studies GRS hulls and intersections
\cite{HuangLiuYu,LiuChen}. Twisted Reed--Solomon codes
\cite{Beelen} provide self-orthogonal and self-dual non-GRS MDS
constructions \cite{Sui,ZhuLiao,ZhuLiaoDouble}.

Polynomial multiplication connects our model with polycyclic
duality \cite{LopezPermouth2009} and with the identity-twist module
codes of Boucher and Ulmer. Over finite fields, when $g(0)\ne0$,
the code is constacyclic or a shortening of a constacyclic code
with the same generator \cite[Proposition~2]{BU2011}.
Shortening retains words vanishing in selected positions and then
deletes those positions. Galois self-orthogonal constacyclic codes
and MDS constructions were studied by Fu and Liu \cite{FuLiu}.
Classifications and GRS representations for cyclic and constacyclic
codes include \cite{RothSeroussi,DahlPedersen,LiuLiu}.
Recent cyclic non-GRS MDS constructions appear in
\cite{LiChenChenNiu,HuangZhangWang}.

For $3\le k\le r$, Li and Yuan showed that, over an algebraic
closure, GRS generators have either distinct roots in a geometric progression or one
repeated root \cite[Theorems~4.4 and~6.6]{LiYuanOrbit}.
These root parameters also determine the dual evaluation nodes
and weights (Lemma~\ref{lem:rootstring}).
That classification and the results on the MDS locus
in \cite{LiYuanSchur} do not impose Euclidean self-orthogonality
in the coefficient coordinates.
We determine how this additional condition restricts the
GRS generators and the MDS codes outside the GRS class.

Theorem~\ref{thm:class} classifies all monic generators giving
self-orthogonal GRS codes over an algebraically closed field.
It gives the precise parameter conditions in both root cases.
For fixed $r,k$ and characteristic, only finitely many generators
occur, and Corollary~\ref{cor:geomcount} gives their exact number.
The classification also determines the largest length for which
each generator gives a self-orthogonal code. At that length the code is cyclic or negacyclic
and GRS. Shorter codes arise by shortening, and every longer code
has distance two (Corollary~\ref{cor:maximal-length}).
These lengths agree with the maximal MDS lengths in
\cite[Theorem~6.13 and Corollary~6.17]{LiYuanSchur}.

For $3\le k<r$, our second main result gives infinitely many
pairwise monomially inequivalent self-orthogonal non-GRS codes
with parameters $[r+k,k,r+1]$ over every algebraically closed field
(Theorem~\ref{thm:exist}).
A uniform bound on the number of monic self-orthogonal MDS
generators in one monomial equivalence class
(Proposition~\ref{prop:monomial-fibers}) allows us to pass from
generators to inequivalent codes.
Consequently, for every finite field $\F_q$ and every positive
integer $B$, some common finite extension of $\F_q$ admits at
least $B$ such codes. They remain non-GRS and pairwise monomially
inequivalent after extension to $\overline{\F}_q$
(Corollary~\ref{cor:abundance}).
At the endpoint $k=r$, only finitely many monic self-dual
generators occur for fixed $r$
(Proposition~\ref{prop:selfdual-endpoint}).

For each finite field $\F_q$, we determine the least extensions
of $\F_q$ containing the coefficients of the classified GRS
generators, and count those already defined over $\F_q$
(Proposition~\ref{prop:arith-orbits} and Theorem~\ref{thm:count}).

Section~\ref{sec:prelim} develops the matrix and evaluation tools.
Section~\ref{sec:class} proves the GRS classification, and
Section~\ref{sec:mds} establishes the non-GRS existence theorem.
Section~\ref{sec:arith} determines coefficient fields and finite-field
counts, and Section~\ref{sec:conclusion} concludes the paper.

\section{Coefficient codes, duality and distance}\label{sec:prelim}

A linear code $C\subseteq K^n$ of dimension $k$ and minimum Hamming
distance $d$ is called an $[n,k,d]$ code. When $K=\F_q$, we write
$[n,k,d]_q$ to specify the field. The Euclidean dual of $C$ is
\[
C^\perp
=\left\{v\in K^n:
\sum_{i=0}^{n-1}v_iw_i=0
\text{ for every }w\in C\right\}.
\]
We say that $C$ is \emph{self-orthogonal} if $C\subseteq C^\perp$,
and \emph{self-dual} if $C=C^\perp$.

\subsection{The shift matrix and the remainder matrix}
Write
\[
g(x)=\sum_{j=0}^{r}g_jx^j,
\]
and set $g_j=0$ for $j\notin\{0,\ldots,r\}$.
The coefficient vectors of $g(x),xg(x),\ldots,x^{k-1}g(x)$
form the rows of a $k\times n$ generator matrix $G$ for $\cC_g(n)$.
Explicitly,
\[
G_{a,i}=g_{i-a},
\qquad 0\le a<k,\quad 0\le i<n.
\]
Thus successive rows shift the coefficients one position to the
right, and $G$ is Toeplitz.

Using the zero-coefficient convention, define for every integer
$d\ge0$
\begin{equation}\label{eq:autocorr}
	R_d(g)=\sum_{j=0}^{r}g_jg_{j+d}.
\end{equation}
The terms with $j+d>r$ vanish. In particular,
\[
R_0(g)=\sum_{j=0}^{r}g_j^2,
\qquad
R_d(g)=0\quad\text{for }d>r.
\]
Two generating rows whose shifts differ by $d$ positions have
inner product $R_d(g)$. This gives the following description
of the Gram matrix $GG^T$.

\begin{lemma}\label{lem:gram}
	For $0\le a,b<k$, we have
	\[
	(GG^T)_{a,b}=R_{|a-b|}(g).
	\]
	Consequently, $\cC_g(n)$ is self-orthogonal if and only if
	\[
	R_0(g)=R_1(g)=\cdots=R_{k-1}(g)=0.
	\]
\end{lemma}

\begin{proof}
For $a\le b$, the zero-coefficient convention gives
\[
(GG^T)_{a,b}
=\sum_{i=0}^{n-1}g_{i-a}g_{i-b}
=\sum_{j=0}^{r}g_{j+b-a}g_j
=R_{b-a}(g).
\]
The other case follows by symmetry. The code is self-orthogonal
exactly when all entries of $GG^T$ vanish.
\end{proof}

Polynomial remainders give the following parity-check matrix
\cite[Proposition~2.4]{LiYuanOrbit}.

\begin{proposition}\label{prop:test}
	For $0\le i<n$, let $Q_i\in K^r$ be the coefficient vector of the
	remainder of $x^i$ upon division by $g$, with respect to the basis
	$1,x,\ldots,x^{r-1}$. Then
	\[
	H=(Q_0\ \cdots\ Q_{n-1})
	\]
	is a parity-check matrix of $\cC_g(n)$, with
	\[
	Hw^T=0\quad\Longleftrightarrow\quad w\in\cC_g(n).
	\]
	Moreover, the following conditions are equivalent:
	\begin{enumerate}
		\item The code $\cC_g(n)$ is MDS.
		\item Every $r$ columns of $H$ are linearly independent.
		\item Every nonzero multiple of $g$ of degree less than $n$ has
		at least $r+1$ nonzero coefficients.
	\end{enumerate}
\end{proposition}

\begin{proof}
For $w=(w_0,\ldots,w_{n-1})$, the vector $Hw^T$ is the remainder
of $\sum_i w_ix^i$ modulo $g$, written in the stated basis.
Thus $\ker H=\cC_g(n)$, and the first $r$ columns of $H$ form
the identity matrix. The equivalence of (1) and (2) is the MDS
parity-check criterion \cite[Theorem~17 and Lemma~18]{BallLavrauw}.
Condition (3) states that the distance is at least $r+1$, which
is equivalent to (1) by the Singleton bound.
\end{proof}

The corresponding generator-matrix criterion states that
$\cC_g(n)$ is MDS if and only if every $k$ columns of $G$
are linearly independent, or equivalently, every $k\times k$
minor of $G$ is nonzero.

Let $T$ be multiplication by $x$ on $K[x]/(g)$ in the basis
$1,x,\ldots,x^{r-1}$. Then $T$ is the companion matrix of $g$,
and its orbit gives the columns of $H$ through
\[
Q_i=T^iQ_0,\qquad 0\le i<n.
\]
When $g(0)\ne0$, $T$ is invertible. We use the resulting
projective orbit in Section~\ref{sec:class}.

\subsection{GRS codes and rational normal curves}

We use the projective definition of GRS codes, which includes
extended GRS codes \cite[Section~13.2.1]{HuffmanPless}.
Choose representatives $(s_i,t_i)$ of $n$ distinct points of
$\PP^1(K)$ and nonzero weights $v_i\in K$. For $1\le a<n$,
the associated dimension-$a$ GRS code is
\[
\bigl\{(v_iF(s_i,t_i))_{i=0}^{n-1}:
F\text{ is homogeneous of degree }a-1\bigr\}.
\]
Here $F$ is homogeneous in two variables. The point $[0:1]$
gives evaluation at infinity, and changing its representative
changes only the corresponding weight. These codes are MDS.

\begin{remark}\label{rem:dimension-range}
	The cases outside $3\le k\le r$ can be described directly.
	If $k>r$, self-orthogonality is impossible because
	$\dim\cC_g(n)=k>r=\dim\cC_g(n)^\perp$.
	The cases $k=1,2$ admit elementary descriptions.
	
	If $k=1$, the code is spanned by $(g_0,\ldots,g_r)$.
	It is self-orthogonal exactly when $R_0(g)=0$, and MDS
	exactly when all coefficients $g_j$ are nonzero.
	In the MDS case, these coefficients can serve as GRS weights
	provided that $\PP^1(K)$ contains at least $n$ distinct points.
	Over $\F_q$, this is equivalent to $n\le q+1$.
	
	If $k=2$, self-orthogonality is equivalent to
	$R_0(g)=R_1(g)=0$, by Lemma~\ref{lem:gram}.
	When the code is MDS, its generator columns are nonzero
	and pairwise nonproportional, giving distinct points of
	$\PP^1(K)$ and hence a GRS representation.
	The corresponding finite-field MDS criteria and GRS
	statements are recorded in
	\cite[Theorem~A.1]{LiYuanOrbit}.
	
	Thus, over an algebraically closed field, every MDS code
	of dimension one or two is GRS. This explains why the
	non-GRS questions considered here begin with $k=3$.
\end{remark}

For $a\ge2$, the standard \emph{rational normal curve} is
\[
[s:t]\longmapsto
[s^{a-1}:s^{a-2}t:\cdots:t^{a-1}]
\quad\text{in }\PP^{a-1}.
\]
Its images under projective changes of coordinates over $K$
are called \emph{split rational normal curves}.
For $a\ge2$, a dimension-$a$ code is GRS if and only if the
projective points represented by its generator columns are
distinct and lie on such a curve.

The dual of a GRS code is again a GRS code on the same evaluation
points, with suitable nonzero weights
\cite[Proposition~2.5]{LiYuanOrbit}.
Consequently, when $r=n-a\ge2$, a dimension-$a$ code is GRS
if and only if the projective points represented by its
parity-check columns are distinct and lie on a split rational
normal curve in $\PP^{r-1}$.

For a linear code $C\subseteq K^n$, its \emph{Schur square} is
\[
C^{\star2}
:=\operatorname{span}\{u\star v:u,v\in C\},
\qquad
u\star v=(u_iv_i)_{i=0}^{n-1}.
\]
For affine points and $2a-1\le n$, the following formula appears
in \cite[Proposition~6]{CouvreurSquare}. For projective points,
products of binary forms of degree $a-1$ span all forms of degree
$2a-2$, whose evaluations at $n$ distinct points have rank
$\min\{n,2a-1\}$. This also covers the range $2a-1>n$.

\begin{lemma}\label{lem:schur-test}
	If $C$ is a dimension-$a$ GRS code of length $n$, including
	the extended case, then
	\[
	\dim C^{\star2}=\min\{n,2a-1\}.
	\]
	Consequently, a dimension-$a$ linear code satisfying
	$\dim C^{\star2}>2a-1$ is not GRS.
\end{lemma}

\begin{example}\label{ex:eleven}
	Over $\F_{11}$, take $r=4$, $k=3$ and
	\[
	g=x^4+6x^3-x^2+4x+1.
	\]
	Then $C=\cC_g(7)$ has generator matrix
	\[
	G=\begin{pmatrix}
		1&4&10&6&1&0&0\\
		0&1&4&10&6&1&0\\
		0&0&1&4&10&6&1
	\end{pmatrix}.
	\]
	A direct calculation gives $GG^T=0$, so $C$ is self-orthogonal.
	Moreover, its $3\times3$ minors take only the nonzero values
	$1,4,6,10$ in $\F_{11}$. Hence $C$ is an $[7,3,5]_{11}$ MDS code.
	
	To show that $C$ is not GRS, we use its Schur square.
	Let $u_0,u_1,u_2$ be the rows of $G$. The six vectors
	$u_a\star u_b$, with $0\le a\le b\le2$, span $C^{\star2}$.
	Arranged in the order
	$(a,b)=(0,0),(0,1),(0,2),(1,1),(1,2),(2,2)$,
	their first six coordinates form a matrix of determinant
	$9\ne0$. Thus $\dim C^{\star2}=6>5$, so $C$ is not GRS
	by Lemma~\ref{lem:schur-test}.
	The same rank calculation holds after extending scalars,
	so $C$ remains non-GRS over the algebraic closure.
\end{example}

\section{The self-orthogonal GRS generators}\label{sec:class}

For the classification, let $K$ be algebraically closed, $3\le k\le r$
and $n=r+k$.

\subsection{Root patterns and dual evaluation codes}

\begin{lemma}\label{lem:rootstring}
	If $\cC_g(n)$ is GRS, one of the following holds.
	\begin{enumerate}
		\item The roots of $g$ are distinct and have the form
		$c,ct,\ldots,ct^{r-1}$ with $c,t\in K^\times$. Put $a_i=t^i$.
		\item We have $g=(x-c)^r$ with $c\in K^\times$, and the
		characteristic is zero or a prime $p\ge n$. Put $a_i=i$.
	\end{enumerate}
	In both cases the $n$ nodes $a_i$ are distinct, and
	\[
	\cC_g(n)^\perp
	=\bigl\{(c^if(a_i))_{i=0}^{n-1}: f\in K[X],\ \deg f<r\bigr\}.
	\]
\end{lemma}

\begin{proof}
	Since $\cC_g(n)$ is GRS, it is MDS and hence $g(0)\ne0$.
	Moreover, $r\ge3$ and $n=r+k\ge r+3$, so the orbit
	classification of Li and Yuan gives the two root patterns
	and the characteristic restriction
	\cite[Theorems~4.4 and~6.6]{LiYuanOrbit}.
	
	In the distinct-root case, a polynomial is divisible by $g$
	if and only if it vanishes at all its roots. Evaluation at these
	roots therefore gives a parity-check matrix whose entries are
	\[
	H_{j,i}=(ct^j)^i=c^i(t^i)^j,
	\qquad 0\le j<r,\quad 0\le i<n.
	\]
	Its row space is the asserted dual code, with nodes $a_i=t^i$
	and weights $c^i$. The nodes are distinct, since equal nodes
	would give proportional columns of $H$, contrary to the
	MDS property.
	
	In the repeated-root case, a polynomial is divisible by
	$(x-c)^r$ if and only if its first $r$ coefficients in the
	expansion at $c$ vanish. For the monomial $x^i$, these coefficients are
	\[
	[Y^j](c+Y)^i=c^{i-j}\binom{i}{j},
	\qquad 0\le j<r.
	\]
	They therefore give a parity-check matrix. Multiplying row $j$
	by $c^j$ changes its entries to $c^i\binom{i}{j}$.
	The characteristic restriction ensures that $j!$ is invertible
	for $0\le j<r$. Thus the polynomials $\binom{X}{j}$ have
	degrees $j$ and form a basis of the polynomials of degree
	less than $r$. Changing to the monomial basis gives entries
	$c^ii^j$, proving the asserted dual representation with $a_i=i$.
	These nodes are distinct because the characteristic is zero
	or at least $n$.
\end{proof}

\subsection{Interpolation and the classification}

Over finite fields the following criterion is equivalent to that of
Fang, Liu and Luo \cite[Lemma~II.2]{FangLiuLuo}.

\begin{lemma}\label{lem:grs-interpolation}
	Let $K$ be any field and $n=r+k$, where $1\le k\le r$.
	Suppose
	\[
	C^\perp=\bigl\{(v_if(a_i))_{i=0}^{n-1}:f\in K[X],\ \deg f<r\bigr\}
	\]
	for distinct $a_i\in K$ and nonzero $v_i\in K$.
	Put
	\[
	D(X)=\prod_{i=0}^{n-1}(X-a_i),\qquad
	w_i=\frac{1}{v_iD'(a_i)}.
	\]
	Then
	\[
	C=\bigl\{(w_ib(a_i))_{i=0}^{n-1}:b\in K[X],\ \deg b<k\bigr\}.
	\]
	The code $C$ is self-orthogonal if and only if some $h\in K[X]$
	satisfies
	\begin{equation}\label{eq:class-interpolation}
		h(a_i)=\frac{w_i}{v_i}=\frac{1}{v_i^2D'(a_i)}
		\quad(0\le i<n),\qquad \deg h\le r-k.
	\end{equation}
\end{lemma}

\begin{proof}
	The formula for $C$ is GRS duality \cite{HuffmanPless}.
	Suppose first that such a polynomial $h$ exists. For every
	$b$ with $\deg b<k$, the bound $\deg h\le r-k$ gives
	$\deg(hb)<r$. Hence
	\[
	(w_ib(a_i))_i=(v_i(hb)(a_i))_i\in C^\perp.
	\]
	Every word of $C$ has this form, proving $C\subseteq C^\perp$.

	Conversely, if $C\subseteq C^\perp$, taking $b=1$ gives a nonzero
	$h$ of degree $d<r$ with $h(a_i)=w_i/v_i$. If $d>r-k$, choose
	$b=X^{r-d}$, so that
	\[
	\deg b=r-d<k,\qquad \deg(hb)=r.
	\]
	Since $\deg b<k$, the word $(w_ib(a_i))_i$ belongs to $C$ and
	therefore to $C^\perp$. Thus some polynomial $f$ with $\deg f<r$
	satisfies
	\[
	v_if(a_i)=w_ib(a_i)=v_ih(a_i)b(a_i)
	\qquad(0\le i<n).
	\]
	All weights are nonzero, so $hb-f$ vanishes at the $n$ nodes.
	Its degree is exactly $r$, because $\deg(hb)=r$ and $\deg f<r$.
	This contradicts $n>r$ and proves $d\le r-k$.
\end{proof}

In both root patterns the nodes are consecutive iterates of an affine
map. Comparing the interpolation values at adjacent nodes gives the
identity used to determine the root parameters.

\begin{lemma}\label{lem:consecutive-nodes}
	Under the hypotheses of Lemma~\ref{lem:grs-interpolation}, assume
	$k\ge2$, $v_i=c^i$ for some $c\in K^\times$, and
	$a_{i+1}=S(a_i)$ for $0\le i<n-1$, where $S(X)=\alpha X+\beta$
	and $\alpha\ne0$. If $C$ is self-orthogonal, there is a nonzero
	$h$ of degree at most $r-k$ such that
	\begin{equation}\label{eq:class-dilation}
		(S(X)-a_0)h(S(X))
		=c^{-2}\alpha^{2-n}(X-a_{n-1})h(X).
	\end{equation}
\end{lemma}

\begin{proof}
	For $h$ from \eqref{eq:class-interpolation} we have
	$D'(a_i)h(a_i)=c^{-2i}$. In the products defining $D'(a_i)$
	and $D'(a_{i+1})$, all but one factor in each product are related
	by $S(a_i)-S(a_j)=\alpha(a_i-a_j)$. Hence
	\[
	\frac{D'(a_{i+1})}{D'(a_i)}
	=\alpha^{n-2}\frac{a_{i+1}-a_0}{a_i-a_{n-1}}
	\qquad(0\le i<n-1).
	\]
	Combining the derivative ratio with the prescribed values of $h$ gives
	\[
	(S(a_i)-a_0)h(S(a_i))
	=c^{-2}\alpha^{2-n}(a_i-a_{n-1})h(a_i).
	\]
	Thus the two sides of \eqref{eq:class-dilation} agree at
	$a_0,\ldots,a_{n-2}$. Their degrees are at most $r-k+1<n-1$,
	so these $n-1$ equalities imply the polynomial identity.
\end{proof}

For the converse, it is convenient to work at the full length of
the multiplicative or additive orbit.

\begin{lemma}\label{lem:full-length}
	Let $K$ be any field and $r<M\le2r$. Suppose that either
	\begin{enumerate}
		\item $g=\prod_{j=0}^{r-1}(x-ct^j)$, where $c,t\in K^\times$,
		$\operatorname{ord}(t)=M$, and $c^2=t^{1-u}$ for an integer
		$0\le u\le2r-M$, or
		\item $M=\operatorname{char} K$ is prime and $g=(x-c)^r$,
		where $c\in K^\times$ and $c^2=1$.
	\end{enumerate}
	Put $\ell=M-r$ and $\varepsilon=c^M$. Then $\cC_g(M)$ is a
	self-orthogonal GRS code and
	\[
	R_j(g)=0\quad(0\le j<\ell),\qquad
	R_\ell(g)=-\varepsilon g(0)\ne0.
	\]
	For every $r<L\le M$, shortening the last $M-L$ positions gives
	the self-orthogonal GRS code $\cC_g(L)$.
\end{lemma}

\begin{proof}
	In case (1), root evaluation gives dual nodes $a_i=t^i$ and
	weights $v_i=c^i$. The nodes are all the $M$th roots of unity.
	Since $t$ has order $M$, the integer $M$ is invertible in $K$, and
	\[
	D(X)=X^M-1,\qquad D'(a_i)=\frac{M}{a_i}.
	\]
	Using $c^2=t^{1-u}$ and $a_i=t^i$, we obtain
	\[
	\frac{1}{v_i^2D'(a_i)}
	=\frac{a_i}{M c^{2i}}=\frac{a_i^u}{M}.
	\]
	Thus $h=X^u/M$ satisfies \eqref{eq:class-interpolation}, and
	the bound on $u$ gives $\deg h\le2r-M=r-\ell$.
	
	In case (2), the Taylor-coefficient calculation above applies
	because $r<M=\operatorname{char}K$. It gives dual nodes $a_i=i$
	and weights $v_i=c^i$. These nodes run through the prime field, so
	\[
	D(X)=X^M-X,\qquad D'(a_i)=-1,\qquad v_i^2=1.
	\]
	The interpolation condition therefore holds with $h=-1$,
	whose degree is at most $r-\ell$.
	Since $1\le\ell=M-r\le r$, Lemma~\ref{lem:grs-interpolation}
	applies with dimension $\ell$ and shows that $\cC_g(M)$
	is GRS and self-orthogonal in both cases.
	
	Since $\cC_g(M)$ is self-orthogonal and has dimension $\ell$,
	Lemma~\ref{lem:gram} gives $R_j(g)=0$ for $0\le j<\ell$.
	In case (1), every root of $g$ has $M$th power $c^M$.
	Since these roots are distinct, $g\mid x^M-c^M$.
	In case (2), the same divisibility follows from
	$x^M-c^M=(x-c)^M$. Thus $g\mid x^M-\varepsilon$.
	The quotient is monic of degree $M-r=\ell$, so we can write
	\[
	x^M-\varepsilon
	=\left(x^\ell+\sum_{j=0}^{\ell-1}b_jx^j\right)g.
	\]
	Regard these polynomials as coefficient vectors of length
	$M+1$, padding $g$ with zeros. Since $M>r$, the inner product
	of $g$ with $x^M-\varepsilon$ is $-\varepsilon g(0)$,
	while its inner product with $x^jg$ is $R_j(g)$. Hence
	\[
	-\varepsilon g(0)
	=R_\ell(g)+\sum_{j=0}^{\ell-1}b_jR_j(g)
	=R_\ell(g)\ne0.
	\]
	This identifies $R_\ell(g)$ as the first nonzero value.
	
	Finally, a word vanishes in the last $M-L$ positions precisely
	when its polynomial has degree less than $L$.
	Shortening in these positions therefore gives $\cC_g(L)$
	and preserves self-orthogonality. Its parity-check matrix is
	obtained by retaining the first $L$ columns of the dual
	evaluation matrix above. The remaining nodes are still distinct,
	so $\cC_g(L)$ is GRS as well.
\end{proof}

\begin{theorem}\label{thm:class}
	Let $K$ be algebraically closed, $3\le k\le r$ and $n=r+k$.
	A monic polynomial $g$ of degree $r$ gives a self-orthogonal
	GRS code $\cC_g(n)$ if and only if one of the following holds.
	\begin{enumerate}
		\item
		\emph{Distinct roots.}
		There are integers
		\[
		n\le m\le2r,\qquad 0\le u\le2r-m,
		\]
		a primitive $m$th root $t\in K$, and $c\in K^\times$ such that
		\[
		g(x)=\prod_{j=0}^{r-1}(x-ct^j),
		\qquad c^2=t^{1-u}.
		\]
		The first nonzero $R_d(g)$ is $R_{m-r}(g)$.
		In characteristic $p>0$, the order $m$ is prime to $p$.
		
		\item
		\emph{One repeated root.}
		The characteristic is a prime $p$ satisfying
		$n\le p\le2r$, and
		\[
		g=(x-1)^r\quad\text{or}\quad g=(x+1)^r.
		\]
		The first nonzero $R_d(g)$ is $R_{p-r}(g)$.
	\end{enumerate}
\end{theorem}

\begin{proof}
	Suppose $\cC_g(n)$ is self-orthogonal and GRS. Apply
	Lemma~\ref{lem:consecutive-nodes} to the two dual representations
	in Lemma~\ref{lem:rootstring}.

	\emph{Distinct roots.}
	With $a_i=t^i$ and $S(X)=tX$, \eqref{eq:class-dilation} reads
	\[
	(tX-1)h(tX)=c^{-2}t^{2-n}(X-t^{n-1})h(X).
	\]
	Let $u$ and $b$ be the smallest and largest exponents occurring
	in $h$, so $0\le u\le b\le r-k$. Comparing the coefficients
	of $X^u$ and $X^{b+1}$ gives
	\[
	t^u=c^{-2}t,\qquad t^{b+1}=c^{-2}t^{2-n}.
	\]
	Thus
	\[
	c^2=t^{1-u},\qquad t^{n+b-u}=1.
	\]
	The order $m$ of $t$ is at least $n$ because the nodes are distinct.
	Since
	\[
	m\mid n+b-u,\qquad n\le n+b-u\le 2r<2m,
	\]
	we have $m=n+b-u$. Consequently $n\le m\le2r$ and
	$0\le u=n+b-m\le2r-m$.
	In characteristic $p>0$, the order $m$ is prime to $p$.

	\emph{One repeated root.}
	With $a_i=i$ and $S(X)=X+1$, the identity becomes
	\[
	(X+1)h(X+1)=c^{-2}(X-n+1)h(X).
	\]
	Write $b=\deg h$. Comparison of leading coefficients gives $c^2=1$.
	Comparison of the coefficients of $X^b$ then gives
	\[
	n+b=0\qquad\text{in }K.
	\]
	The positive integer $n+b$ therefore vanishes in $K$, so the
	characteristic is a prime $p$ dividing $n+b$. The root-pattern
	lemma gives $p\ge n$, whence
	\[
	p\mid n+b,\qquad n\le n+b\le2r<2p.
	\]
	The only possible positive multiple is $n+b=p$. Thus
	$n\le p\le2r$ and $c=\pm1$.

	Conversely, set $M=m$ in the distinct-root case and $M=p$ in the
	repeated-root case. Since $n\le M$, Lemma~\ref{lem:full-length}
	gives self-orthogonality and GRS type at length $n$, as well as
	the first nonzero value $R_{M-r}(g)$.
\end{proof}

\begin{remark}\label{rem:mds-existence}
	Let $K$ be algebraically closed and $r\ge2$.
	Take $m=2r$, replacing it by $2r-1$ when the positive
	characteristic divides $2r$. Choose $t\in K^\times$
	of order $m$ and $c\in K^\times$ with $c^2=t$, and set
	\[
	g(x)=\prod_{j=0}^{r-1}(x-ct^j).
	\]
	For every $1\le k<r$, this gives a self-orthogonal GRS
	code $\cC_g(r+k)$ with parameters $[r+k,k,r+1]$.
	Indeed, $r+k\le2r-1\le m$, so Lemma~\ref{lem:full-length}
	applies with $u=0$, including when $k=1,2$.
	If $m=2r$, the same lemma also gives a self-dual GRS code
	$\cC_g(2r)$ with parameters $[2r,r,r+1]$.
\end{remark}

\subsection{Maximal length and generator counts}

We now fix $g$ and vary the code length. The next corollary
determines the largest length for which $\cC_g(L)$ is
self-orthogonal. At that length the code is cyclic or negacyclic.
The length is determined by the first nonzero value among the $R_d(g)$.

\begin{corollary}\label{cor:maximal-length}
	Let $g$ be as in Theorem~\ref{thm:class}, and put
	\[
	\ell=\min\{d\ge0:R_d(g)\ne0\},
	\qquad M=r+\ell.
	\]
	Thus $M=m$ in the distinct-root case, with $c$ as in
	Theorem~\ref{thm:class}. In the repeated-root case,
	$M=p$ and we write $g=(x-c)^r$ with $c\in\{1,-1\}$.
	In both cases, put $\varepsilon=c^M\in\{1,-1\}$.
	Then the following hold.
	\begin{enumerate}
		\item
		The integer $M$ is the least positive $L$ such that
		$g\mid x^L-\lambda$ for some $\lambda\in K^\times$.
		At length $M$, the code $\cC_g(M)$ is the ideal $(g)$
		in $K[x]/(x^M-\varepsilon)$ and is therefore cyclic
		or negacyclic.
		
		\item
		For every $r<L\le M$, the code $\cC_g(L)$ is a
		self-orthogonal GRS code with parameters
		\[
		[L,L-r,r+1].
		\]
		It is obtained by shortening $\cC_g(M)$ in its last
		$M-L$ coordinate positions.
		
		\item
		For every $L>M$, the code $\cC_g(L)$ is not
		self-orthogonal and has minimum distance exactly two.
	\end{enumerate}
\end{corollary}

\begin{proof}
	Lemma~\ref{lem:full-length} and its proof give
	$g\mid x^M-\varepsilon$ and show that $\cC_g(M)$ is
	self-orthogonal and GRS. The multiples of $g$ of degree
	less than $M$ identify this code with the ideal $(g)$ in
	$K[x]/(x^M-\varepsilon)$. In the distinct-root case,
	$\varepsilon^2=c^{2M}=t^{M(1-u)}=1$; in the repeated-root
	case this follows from $c^2=1$. Thus $\varepsilon\in\{1,-1\}$,
	and the code is cyclic or negacyclic.
	
	To prove minimality, suppose that $g\mid x^L-\lambda$
	for some positive $L$ and nonzero $\lambda$.
	In the distinct-root case, evaluation at the roots $c$ and
	$ct$ gives
	\[
	c^L=(ct)^L=\lambda,
	\]
	so $t^L=1$ and hence $m\mid L$.
	In the repeated-root case, $c$ is a multiple root of
	$x^L-\lambda$. Its derivative therefore vanishes at $c$,
	giving $Lc^{L-1}=0$. Since $c\ne0$, this implies $p\mid L$.
	Thus $M=m$ or $M=p$, respectively, is the least possible length.
	
	For $r<L\le M$, Lemma~\ref{lem:full-length} gives the
	shortened self-orthogonal GRS code $\cC_g(L)$.
	Its dimension $L-r$ gives minimum distance $r+1$.
	
	Finally, let $L>M$. Since $g\mid x^M-\varepsilon$ and
	$M<L$, the polynomial $x^M-\varepsilon$ gives a codeword
	of weight two in $\cC_g(L)$. Its inner product with the
	codeword $g$ is
	\[
	\langle g,x^M-\varepsilon\rangle
	=-\varepsilon g(0)\ne0,
	\]
	where we use $M>r$. Thus $\cC_g(L)$ is not self-orthogonal.
	Moreover, $g(0)\ne0$ and $\deg g>0$ ensure that no nonzero
	monomial is divisible by $g$. Hence the code has no word
	of weight one, and its minimum distance is exactly two.
\end{proof}

Over finite fields, Boucher and Ulmer realize module codes with
$g(0)\ne0$ as constacyclic codes or their shortenings in the
identity-twist case \cite[Proposition~2]{BU2011}.
The shortening rule for self-orthogonal MDS codes is also known
\cite[Corollary~2.13]{LiSu}, as is the GRS property of length-$p$
constacyclic codes in characteristic $p$ \cite{PedersenDahl}.
For the generators classified here, Corollary~\ref{cor:maximal-length}
determines the least positive $L$ for which $g\mid x^L-\lambda$
for some $\lambda\in K^\times$ and shows that it equals the
maximal self-orthogonal length.

\begin{corollary}\label{cor:geomcount}
	Let $3\le k\le r$, put $n=r+k$, and let $A_{r,k}(p)$
	count the distinct monic generators in
	Theorem~\ref{thm:class} over an algebraically closed field
	of characteristic $p$, allowing $p=0$.
	Let $\varphi$ denote Euler's totient function, and set
	$\mathbf 1_E=1$ when $E$ holds and $0$ otherwise. Then
	\begin{align*}
		A_{r,k}(0)
		&=\sum_{m=n}^{2r}\varphi(m)(2r-m+1),\\
		A_{r,k}(p)
		&=\sum_{\substack{n\le m\le2r\\p\nmid m}}
		\varphi(m)(2r-m+1)
		+2\,\mathbf 1_{\{n\le p\le2r\}}
		&&(p\text{ odd}),\\
		A_{r,k}(2)
		&=\frac12\sum_{\substack{n\le m\le2r\\m\text{ odd}}}
		\varphi(m)(2r-m+1).
	\end{align*}
\end{corollary}

\begin{proof}
	First consider the distinct-root case. Fix an admissible order
	$n\le m\le2r$ and put $D=2r-m$. Since $n=r+k$ and $k\ge3$,
	we have $0\le D\le r-k\le r-3$.
	
	Listing the roots in reverse order gives the same polynomial,
	with first root $ct^{r-1}$ and ratio $t^{-1}$.
	The corresponding parameter change is
	\begin{equation}\label{eq:reversal}
		(u,t,c)\longmapsto(D-u,t^{-1},ct^{r-1}).
	\end{equation}
	Since $0\le u\le D$, the new integer $D-u$ also lies between
	$0$ and $D$. Moreover, using $c^2=t^{1-u}$ and $t^m=1$, we obtain
	\[
	(ct^{r-1})^2
	=t^{2r-1-u}
	=(t^{-1})^{1-(D-u)}.
	\]
	Thus the new triple is admissible and represents the same $g$.
	It differs from the original triple because $m\ge n\ge6$
	ensures that $t\ne t^{-1}$.
	
	We next show that these are the only two representations
	of the same polynomial. Let
	$S=\{c,ct,\ldots,ct^{r-1}\}$ be its root set.
	The quotients of elements of $S$ generate exactly
	$\langle t\rangle$, since they are powers of $t$ and include
	$t$ itself. Thus $S$ determines $m$.
	For $1\le a<m$, the number of ordered pairs
	$(\alpha,\beta)\in S^2$ with $\beta/\alpha=t^a$ is
	\[
	\max\{r-a,0\}+\max\{r-m+a,0\}.
	\]
	Indeed, writing $\alpha=ct^i$ and $\beta=ct^j$, the possible
	index differences are $j-i=a$ and $j-i=a-m$.
	When both summands are positive, their sum is
	$D\le r-3$. When only one is positive, the value $r-1$
	occurs exactly at $a=1$ and $a=m-1$.
	
	Any progression representing $S$ has $r-1$ consecutive pairs
	with the same quotient. Its ratio must therefore be $t$
	or $t^{-1}$. For a fixed ratio, the first root is the unique
	element of $S$ whose predecessor is absent from $S$.
	This determines $c$, and $c^2=t^{1-u}$ then determines $u$,
	since $0\le u\le D<m$. Hence each polynomial is represented
	by exactly the two triples in \eqref{eq:reversal}.
	
	For fixed $m$, there are $\varphi(m)$ choices for $t$ and
	$D+1$ choices for $u$. The equation $c^2=t^{1-u}$ has two
	solutions in characteristic different from two and one in
	characteristic two. Dividing the number of triples by two
	therefore gives the stated contribution for each $m$.
	In characteristic zero every order in $[n,2r]$ is admissible.
	In characteristic $p>0$, precisely the orders prime to $p$
	are admissible.
	
	Finally, the repeated-root case contributes the two distinct
	polynomials $(x-1)^r$ and $(x+1)^r$ exactly when
	$n\le p\le2r$. This condition forces $p$ to be odd,
	completing all three formulas.
\end{proof}

\begin{example}\label{ex:count-four-three}
	For $(r,k)=(4,3)$, we have $n=7$, so the possible orders
	are $7$ and $8$. Their contributions in characteristic zero
	are $2\varphi(7)=12$ and $\varphi(8)=4$, giving
	$A_{4,3}(0)=16$.
	In characteristic seven, only order $8$ contributes to
	the distinct-root case, giving four generators.
	The repeated-root generators $(x-1)^4$ and $(x+1)^4$
	bring the total to $A_{4,3}(7)=6$.
	
	To illustrate the repeated-root case, take
	\[
	g=(x-1)^4=x^4+3x^3+6x^2+3x+1
	\qquad\text{over }\F_7.
	\]
	A direct calculation gives
	\[
	R_0(g)=R_1(g)=R_2(g)=0,\qquad R_3(g)=6\ne0.
	\]
	By Lemma~\ref{lem:full-length} with $M=7$, the code $\cC_g(7)$ is
	a self-orthogonal GRS code with parameters $[7,3,5]_7$.
	Its largest self-orthogonal length is $7$.
	At length $8$, the multiple $x^7-1=(x-1)^7$ gives
	a word of weight two, illustrating
	Corollary~\ref{cor:maximal-length}.
\end{example}

The counts above allow coefficients throughout the algebraically
closed field. Section~\ref{sec:arith} determines which generators
are defined over a specified finite field.

\section{Infinitely many non-GRS MDS codes}\label{sec:mds}

Li and Yuan proved density of non-GRS generators on the strata
defined by the first length at which the MDS property fails
\cite[Corollary~6.19]{LiYuanSchur}.
Corollary~\ref{cor:geomcount} gives only finitely many monic
self-orthogonal GRS generators for fixed $r,k$.
We now show that, for $3\le k<r$, the self-orthogonal MDS codes
outside the GRS class represent infinitely many monomial
equivalence classes over every algebraically closed field.
We first bound the number of monic
generators in one equivalence class.

\begin{proposition}\label{prop:monomial-fibers}
	Let $K$ be any field and let $n=r+k$ with $2\le k\le r$.
	A monomial equivalence class contains at most $2(r-k+1)n!$
	distinct monic degree-$r$ generators $g\in K[x]$ for which
	$\cC_g(n)$ is self-orthogonal and MDS.
\end{proposition}

\begin{proof}
	Suppose that a diagonal transformation with entries
	$d_0,\ldots,d_{n-1}$ takes $\cC_g(n)$ to $\cC_h(n)$.
	Since $g$ and $h$ represent codewords of weight at least
	$r+1$, all their coefficients are nonzero.
	For $0\le j<k$, the codewords of $\cC_g(n)$ supported in
	$\{j,\ldots,j+r\}$ are precisely the scalar multiples of $x^jg$.
	Indeed, a nonzero multiple $ug$ with this support satisfies
	$x^j\mid u$ because $g(0)\ne0$, and $\deg u\le j$
	by the degree bound. The same argument applies to $h$.
	
	A diagonal transformation preserves supports. It must therefore
	send each $x^jg$ to a scalar multiple $A_jx^jh$, with
	$A_j\in K^\times$. In coordinate $j+s$ this says
	\[
	d_{j+s}g_s=A_jh_s
	\qquad(0\le j<k,\ 0\le s\le r).
	\]
	Comparing the equations for $j+1$ and $j$ gives
	\[
	\frac{d_{j+s+1}}{d_{j+s}}=\frac{A_{j+1}}{A_j}
	\qquad(0\le j<k-1,\ 0\le s\le r).
	\]
	The intervals $\{j,\ldots,j+r\}$ overlap and cover
	$\{0,\ldots,n-2\}$. Hence all adjacent ratios are equal
	to some $t\in K^\times$, and $d_i=d_0t^i$.
	Using $g_r=h_r=1$ in the coordinate equations now gives
	\[
	h_s=t^{s-r}g_s\quad(0\le s\le r),
	\qquad h(x)=t^{-r}g(tx).
	\]
	Conversely, this dilation gives a diagonal equivalence
	\cite[Remark~4.7]{LiYuanSchur}.
	
	The equation $R_{k-1}(h)=0$ implies
	\[
	\sum_{s=0}^{r-k+1}g_sg_{s+k-1}t^{2s}=0.
	\]
	Its leading coefficient $g_{r-k+1}g_r$ is nonzero, so
	$t$, and hence $h$, has at most $2(r-k+1)$ possibilities.
	For each permutation of the coordinates of a fixed code,
	the generators obtainable by subsequent diagonal transformations
	belong to one diagonal equivalence class.
	Summing over the $n!$ permutations gives the asserted bound.
\end{proof}

\begin{theorem}\label{thm:exist}
	Let $K$ be algebraically closed, $3\le k<r$, and $n=r+k$.
	There are infinitely many pairwise monomially inequivalent
	self-orthogonal non-GRS codes $\cC_g(n)$ with parameters
	$[n,k,r+1]$, generated by monic degree-$r$ polynomials
	$g\in K[x]$.
\end{theorem}

\begin{proof}
	Write
	\[
	g(x)=g_0+g_1x+\cdots+g_{r-1}x^{r-1}+x^r,
	\]
	so that $g_0,\ldots,g_{r-1}$ are the $r$ unknown coefficients.
	Let $G_I(g)$ be the submatrix of the generator matrix $G$
	formed by the columns indexed by $I$, and put
	\[
	\Delta(g)=
	\prod_{\substack{I\subseteq\{0,\ldots,n-1\}\\|I|=k}}
	\det G_I(g).
	\]
	The generator-matrix criterion in Section~\ref{sec:prelim}
	shows that $\Delta(g)\ne0$ precisely when $\cC_g(n)$ is MDS.
	Introducing one additional unknown $z$ turns this nonvanishing
	condition into the equation $z\Delta(g)=1$.
	Thus, by Lemma~\ref{lem:gram}, the solutions of
	\begin{equation}\label{eq:mds-system}
		R_0(g)=\cdots=R_{k-1}(g)=0,
		\qquad z\Delta(g)-1=0
	\end{equation}
	correspond exactly to the monic self-orthogonal MDS generators,
	with $z=\Delta(g)^{-1}$ uniquely determined by $g$.
	Remark~\ref{rem:mds-existence} supplies one such generator,
	so the system has a solution.
	
	Let $X\subseteq K^{r+1}$ be the affine algebraic set defined
	by~\eqref{eq:mds-system}, with coordinates
	$g_0,\ldots,g_{r-1},z$.
	Since $X$ is nonempty and is defined by $k+1$ polynomial
	equations in $r+1$ variables, Krull's height theorem implies
	that every irreducible component of $X$ has dimension at least
	\[
	\dim X\ge (r+1)-(k+1)=r-k\ge1;
	\]
	see \cite[Tags~00FV, 00OP and 0BBZ]{Stacks}.
	Hence $X$ has infinitely many $K$-rational points because
	$K$ is algebraically closed.
	
	Moreover, the last equation in~\eqref{eq:mds-system}
	uniquely determines $z$ from $g$.
	Consequently, distinct points of $X$ correspond to distinct
	monic degree-$r$ generators.
	It follows that there are infinitely many monic
	self-orthogonal MDS generators.
	
	Only finitely many of these generators are GRS, by
	Corollary~\ref{cor:geomcount}.
	Removing them leaves infinitely many non-GRS generators.
	Proposition~\ref{prop:monomial-fibers}
	bounds their number in each monomial equivalence class by
	$2(r-k+1)n!$, so infinitely many classes occur.
\end{proof}

The restriction $k<r$ in Theorem~\ref{thm:exist} is essential.
For fixed $r$, only finitely many monic self-orthogonal
generators occur at the endpoint $k=r$.

Over finite fields, the reciprocal criterion below is the
identity-twist specialization of Boucher and Ulmer's
self-duality criterion \cite{BU2014}.
The nonexistence in characteristics dividing $2r$ also follows
from Pedersen and Dahl's repeated-root MDS obstruction
\cite{PedersenDahl}. We include a direct proof.

\begin{proposition}
	\label{prop:selfdual-endpoint}
	Let $K$ be a field and $r\ge3$. A monic polynomial
	$g\in K[x]$ of degree $r$ gives a self-dual code
	$\cC_g(2r)$ if and only if
	\begin{equation}\label{eq:selfdual-factorization}
		g(x)x^rg(x^{-1})=g(0)(x^{2r}+1).
	\end{equation}
	In this case, $g(0)\ne0$ and $g\mid x^{2r}+1$.
	Consequently, for fixed $r$, only finitely many monic
	self-dual generators occur.
	
	If $K$ is algebraically closed, a self-dual MDS code
	$\cC_g(2r)$ exists precisely in characteristic zero
	or in characteristic $p>0$ with $p\nmid2r$.
\end{proposition}

\begin{proof}
	Since $\cC_g(2r)$ has dimension $r$, self-duality is
	equivalent to self-orthogonality.
	The coefficients of $x^{r-d}$ and $x^{r+d}$ in
	$g(x)x^rg(x^{-1})$ are $R_d(g)$ for $0\le d\le r$,
	with $R_r(g)=g(0)$.
	Lemma~\ref{lem:gram} therefore gives
	\eqref{eq:selfdual-factorization} and its converse.
	The product is nonzero, so the identity forces
	$g(0)\ne0$ and $g\mid x^{2r}+1$.
	This fixed polynomial has only finitely many monic divisors.
	
	Now assume that $K$ is algebraically closed.
	In characteristic zero or characteristic prime to $2r$,
	Remark~\ref{rem:mds-existence} with $m=2r$ supplies
	a self-dual GRS code, proving existence.
	
	Suppose instead that $\operatorname{char}K=p\mid2r$,
	and let $\cC_g(2r)$ be self-dual.
	Write $2r=eD$, where $e=p^a\ge p$ and $p\nmid D$.
	Then
	\[
	x^{2r}+1=(x^D+1)^e,
	\]
	and $x^D+1$ has $D$ distinct roots.
	If $D=1$, then $2r=p^a$ is even, so $p=2$ and $r=2^{a-1}$.
	Since $g$ is monic of degree $r$ and divides
	$x^{2r}+1=(x+1)^{2r}$, we have
	$g=(x+1)^r=x^r+1$ by Frobenius.
	Thus $g$ represents a codeword of weight two, below the
	MDS distance $r+1$.
	
	For a polynomial $f$, write $\wt(f)$ for the number of its
	nonzero coefficients.
	Assume $D\ge2$, so both $e$ and $D$ are at most $r$.
	By \eqref{eq:selfdual-factorization}, the multiplicities
	in $g$ of a root and its reciprocal sum to $e$.
	If a root $\alpha$ has multiplicity $e$ in $g$, then
	$\alpha^{-1}$ is one of the $D$ roots of $x^D+1$ and is
	absent from $g$. Hence $g$ has at most $D-1$ distinct roots.
	Let $h$ be the product of the distinct monic linear factors
	of $g$. Then $\deg h\le D-1$ and $g\mid h^e$, while
	\[
	\deg(h^e)<2r,
	\qquad
	\wt(h^e)=\wt(h)\le D\le r,
	\]
	where the weight equality follows from Frobenius.
	Otherwise every root of $g$ has multiplicity at most $e-1$,
	so $g$ divides $(x^D+1)^{e-1}$, which has degree less
	than $2r$ and weight at most $e\le r$.
	In either case, a nonzero multiple of $g$ violates the
	MDS criterion in Proposition~\ref{prop:test}.
\end{proof}

\begin{example}\label{ex:explicit-family}
	Let $K=\overline{\F}_{17}$ and take $r=5$, $k=3$.
	Put
	\[
	h(x)=\prod_{j=0}^{3}(x-3\cdot9^j)
	=x^4+5x^3+4x^2-5x+1,
	\qquad
	g_s(x)=(x-s)h(x).
	\]
	For every $s\in K$ satisfying
	\[
	[\F_{17}(s):\F_{17}]>3,
	\]
	the code $\cC_{g_s}(8)$ is a self-orthogonal non-GRS
	MDS code with parameters $[8,3,6]$.
	
	Indeed, $9=3^2$ has order $8$, so
	Theorem~\ref{thm:class} shows that $\cC_h(8)$ is
	self-orthogonal and MDS. Its dimension is $4$, hence
	it is self-dual.
	Since $\cC_{g_s}(8)\subseteq\cC_h(8)$, every code
	in this family is self-orthogonal.
	At $s=-3=3\cdot9^4$, the five roots of $g_s$ form
	a geometric progression, and Theorem~\ref{thm:class}
	shows that $\cC_{g_{-3}}(8)$ is MDS.
	Every entry of the shift generator matrix is affine in $s$,
	so each $3\times3$ minor has degree at most $3$ over $\F_{17}$.
	Each is nonzero at $s=-3$, hence is a nonzero polynomial.
	None can vanish when $[\F_{17}(s):\F_{17}]>3$,
	proving the MDS assertion.
	
	If $\cC_{g_s}(8)$ were GRS, Theorem~\ref{thm:class}
	would give a root ratio of order $m$ with $8\le m\le10$.
	The quotient of the roots $3\cdot9$ and $3$ is $9$,
	whose order is $8$. Since every quotient of roots is a power
	of the progression ratio, $8\mid m$ and hence $m=8$.
	The root set would then be contained in
	\[
	\{3\cdot9^j:0\le j<8\}\subseteq\F_{17},
	\]
	contradicting $s\notin\F_{17}$.
	
	There are infinitely many such parameters $s$, and
	$g_s(0)=-s$ makes their generators distinct.
	Proposition~\ref{prop:monomial-fibers} therefore shows
	that this family represents infinitely many monomial
	equivalence classes.
\end{example}

Applying Theorem~\ref{thm:exist} over $\overline{\F}_q$ gives
the following consequence for every prime power $q$.

\begin{corollary}\label{cor:abundance}
	Let $q$ be a prime power, $3\le k<r$, and $B\ge1$ an integer.
	There is an integer $m\ge1$ and monic degree-$r$ polynomials
	$g_1,\ldots,g_B\in\F_{q^m}[x]$ such that the codes
	$\cC_{g_i}(r+k)$ are self-orthogonal MDS codes with parameters
	$[r+k,k,r+1]$. After extending scalars to $\overline{\F}_q$,
	they remain non-GRS and pairwise monomially inequivalent.
\end{corollary}

\begin{proof}
	Apply Theorem~\ref{thm:exist} over $\overline{\F}_q$ and
	choose $B$ generators whose codes are pairwise monomially
	inequivalent. Their finitely many coefficients lie in a common
	finite extension $\F_{q^m}$.
	By Lemma~\ref{lem:gram}, the resulting codes are
	self-orthogonal over this field. Their maximal generator
	minors remain nonzero, so they are MDS as well.
	Their scalar extensions are the chosen codes, giving
	the remaining assertions.
\end{proof}

\section{Coefficient fields and finite-field counts}\label{sec:arith}

Let $q$ be a power of a prime $p$, fix $3\le k\le r$, and put $n=r+k$.
Let $N_{r,k}(q)$ count the monic degree-$r$ polynomials $g\in\F_q[x]$
for which $\cC_g(n)$ is self-orthogonal and GRS over $\F_q$.
Self-orthogonality descends to the coefficient field by
Lemma~\ref{lem:gram}, as does GRS type in this range
\cite[Corollary~4.5]{LiYuanOrbit}.
Thus we need to identify the generators of Theorem~\ref{thm:class}
whose coefficients lie in $\F_q$. This holds precisely when
Frobenius $z\mapsto z^q$ preserves their roots with multiplicities.
The roots themselves may lie in an extension of $\F_q$.
The resulting GRS representations have nodes in $\PP^1(\F_q)$
and weights in $\F_q^\times$. If $n\le q$, sending an unused
point to infinity makes all nodes affine.

For $p\nmid s$, write $\operatorname{ord}_s(q)$ for the multiplicative
order of $q$ modulo $s$. For even $m$ prime to $p$, put
\[
a_*(m,q)=\min\{h\ge1:q^h\equiv1\text{ or }m-1\pmod{2m}\}.
\]

\begin{proposition}\label{prop:arith-orbits}
	Fix $n\le m\le2r$ with $p\nmid m$, and put $D=2r-m$ and
	$\eta=\gcd(2,q-1)$. Consider
	\[
	g(x)=\prod_{j=0}^{r-1}(x-ct^j),\qquad
	t,c\in\overline{\F}_q^\times,\quad
	\operatorname{ord}(t)=m,\quad
	0\le u\le D,\quad c^2=t^{1-u}.
	\]
	Table~\ref{tab:coeff-field-degrees} counts the distinct monic polynomials in each parameter
	range. For each polynomial in a nonempty row, the last column gives
	the least integer $a\ge1$ such that $g\in\F_{q^a}[x]$.
	\begin{center}
	\captionof{table}{Generator counts and least coefficient-field degrees.}\label{tab:coeff-field-degrees}
	\smallskip
	$\displaystyle
	\begin{array}{lcc}
		\toprule
		\text{parameters}
		&\text{number of generators}
		&\text{least degree }a\\
		\midrule
		m\text{ odd}
		&\varphi(m)\eta(D+1)/2
		&\operatorname{ord}_m(q)\\
		m\text{ even},\ u\text{ odd}
		&\varphi(m)D/2
		&\operatorname{ord}_m(q)\\
		m,u\text{ even},\ 2u\ne D
		&2\varphi(m)\lceil D/4\rceil
		&\operatorname{ord}_{2m}(q)\\
		m,u\text{ even},\ 2u=D
		&\varphi(m)\mathbf1_{\{4\mid D\}}
		&a_*(m,q)\\
		\bottomrule
	\end{array}$
	\end{center}
	The repeated-root generators $(x\pm1)^r$, when present,
	lie in $\F_p[x]$.
\end{proposition}

\begin{proof}
	For $Q=q^h$, the $Q$-power Frobenius sends $(u,t,c)$
	to $(u,t^Q,c^Q)$.
	The coefficients of $g$ lie in $\F_Q$ exactly when this new
	triple describes the same root set.
	By Corollary~\ref{cor:geomcount}, that set has just two
	descriptions, the original one and its reversal
	$(D-u,t^{-1},ct^{r-1})$ in \eqref{eq:reversal}.
	Hence there are two possibilities:
	\[
	\begin{array}{ll}
		\text{same order:}
		&t^Q=t,\quad c^Q=c,\\
		\text{reverse order:}
		&2u=D,\quad t^Q=t^{-1},\quad c^Q=ct^{r-1}.
	\end{array}
	\]
	If $m$ is odd, then $D$ is odd, so the second possibility
	is excluded.
	Squaring is bijective on $\langle t\rangle$, so $c$ is a
	power of $t$ up to sign.
	Fixing $t$ therefore suffices, giving
	$a=\operatorname{ord}_m(q)$.
	
	Now let $m$ be even. The characteristic is odd, and
	\[
	c^{2m}=1,\qquad
	c^m=(t^{m/2})^{1-u}=(-1)^{1-u}.
	\]
	Thus $c\in\langle t\rangle$ exactly when $u$ is odd.
	Fixing both $t$ and $c$ is equivalent to
	\[
	Q\equiv1\pmod m\quad(u\text{ odd}),\qquad
	Q\equiv1\pmod{2m}\quad(u\text{ even}).
	\]
	For the reverse order, we need $2u=D$ and $m\mid Q+1$.
	Here $r-u=m/2$, so $c^2t^{r-1}=-1$.
	The remaining condition is equivalently
	\[
	c^{Q+1}=(c^m)^{(Q+1)/m}=-1.
	\]
	This is impossible for odd $u$.
	For even $u$, it says that $(Q+1)/m$ is odd, or
	$Q\equiv m-1\pmod{2m}$.
	Together with the same-order case, this gives $a_*(m,q)$.
	
	To count the polynomials, each $u$ gives $\varphi(m)$
	choices of $t$ and $\eta$ choices of $c$.
	Every polynomial is counted twice, once in each direction,
	and reversal preserves each row.
	For odd $m$ there are $D+1$ values of $u$.
	For even $m$ there are $D/2$ odd values, one central even
	value if $4\mid D$, and $2\lceil D/4\rceil$ noncentral
	even values.
	Dividing the triple counts by two gives Table~\ref{tab:coeff-field-degrees}.
\end{proof}

Taking the rows with $a=1$ and adding the repeated-root generators
gives the exact count over $\F_q$.

\begin{theorem}\label{thm:count}
	Let $q$ be a power of a prime $p$, with $3\le k\le r$
	and $n=r+k$. Then
	\[
	N_{r,k}(q)
	=\sum_{\substack{n\le m\le2r\\p\nmid m}}
	\varphi(m)w_q(m)
	+2\,\mathbf1_{\{n\le p\le2r\}}.
	\]
	For $D=2r-m$, the weights $w_q(m)$ are given in
	Table~\ref{tab:finite-field-weights} and are zero in all unlisted cases.
	\begin{center}
	\captionof{table}{Weights for the finite-field generator-count formula.}\label{tab:finite-field-weights}
	\smallskip
	$\displaystyle
	\begin{array}{clc}
		\toprule
		q&\text{condition}&w_q(m)\\
		\midrule
		\text{odd}
		&2m\mid q-1
		&D+1\\
		\text{odd}
		&m\text{ even},\ m\mid q-1,\ 2m\nmid q-1
		&D/2\\
		\text{odd}
		&m\text{ even},\ 4\mid D,\ q\equiv m-1\pmod{2m}
		&1\\
		\text{even}
		&m\text{ odd},\ m\mid q-1
		&(D+1)/2\\
		\bottomrule
	\end{array}$
	\end{center}
\end{theorem}

\begin{proof}
	Sum the counts in Proposition~\ref{prop:arith-orbits}
	with $a=1$.
	For odd $q$ and odd $m$, the conditions $m\mid q-1$
	and $2m\mid q-1$ coincide.
	Theorem~\ref{thm:class} adds the two repeated-root
	generators exactly when $n\le p\le2r$.
\end{proof}

\begin{example}\label{ex:finite-field-counts}
	Take $(r,k)=(4,3)$, so $n=7$ and the possible orders
	are $7$ and $8$.
	Over $\F_{11}$, neither order contributes to
	Theorem~\ref{thm:count}, and $p=11>2r$ excludes
	the repeated-root case.
	Thus $N_{4,3}(11)=0$, so every self-orthogonal MDS
	code $\cC_g(7)$ over $\F_{11}$ is non-GRS,
	including the code in Example~\ref{ex:eleven}.
	
	In characteristic two, only $m=7$ is admissible,
	and no repeated-root generator occurs.
	Applying Proposition~\ref{prop:arith-orbits} with $q=2$
	gives six generators, all with least coefficient field
	$\F_8$, since $\operatorname{ord}_7(2)=3$.
	Consequently, for every $e\ge1$,
	\[
	N_{4,3}(2^e)=
	\begin{cases}
		6,&3\mid e,\\
		0,&3\nmid e.
	\end{cases}
	\]
	In particular, none is defined over $\F_2$ or $\F_4$,
	whereas all six are defined over $\F_8$.
\end{example}

\begin{corollary}\label{cor:arith-endpoints}
	Under the assumptions of Theorem~\ref{thm:count},
	$N_{r,k}(q)=0$ if $n>q+1$, and already if $n>q-1$
	for even $q$. At the two endpoints,
	\begin{align*}
		n=q+1:\qquad N_{r,k}(q)
		&=
		\begin{cases}
			\varphi(q+1),&4\mid r-k,\\
			0,&\text{otherwise},
		\end{cases}\\
		k=r:\qquad N_{r,r}(q)
		&=
		\begin{cases}
			\varphi(2r),
			&q\equiv1\text{ or }2r-1\pmod{4r},\\
			0,&\text{otherwise}.
		\end{cases}
	\end{align*}
\end{corollary}

\begin{proof}
	For a progression defined over $\F_q$, the same-order
	case requires $m\mid q-1$, so $n\le m\le q-1$.
	The reverse-order case requires even $m$ with $m\mid q+1$,
	so $n\le m\le q+1$.
	Repeated roots require $n\le p\le q$ with $p$ odd.
	These observations give both length bounds.
	
	If $n=q+1$, only reversal can contribute, with $m=q+1$
	and $4\mid D=r-k$.
	For even $q$, $r-k$ is odd, so the count is zero.
	If $k=r$, only $m=2r$ is possible and $D=0$.
	The repeated-root term vanishes because $2r\ge6$ is composite.
	Table~\ref{tab:finite-field-weights} now gives both formulas.
\end{proof}

\begin{example}\label{ex:arith-seven}
	For $q=7$ and $k=r=4$, reversal at $m=8$ gives
	$N_{4,4}(7)=4$.
	This illustrates how coefficients can lie in $\F_7$
	while roots lie in $\F_{49}$.
	Take a root $c$ of $x^2-x-1$ in $\F_{49}$.
	Its discriminant $5$ is a nonsquare in $\F_7$, so
	$c^7=1-c$ and $c^8=c(1-c)=-1$.
	Thus $c$ has order $16$, and Frobenius reverses
	$c,c^3,c^5,c^7$. The generator is
	\[
	g(x)=\prod_{j=0}^3(x-c^{2j+1})
	=x^4+2x^3+2x^2+5x+1.
	\]
	The self-dual $[8,4,5]_7$ GRS code $\cC_g(8)$ uses
	all eight points of $\PP^1(\F_7)$.
	It belongs to Blackford's length-$(q+1)$ negacyclic
	construction \cite[Corollary~5]{Blackford} and is the
	$\F_7$-subfield subcode of a GRS code over $\F_{49}$
	\cite[Proposition~5.9(ii)--(iii) in the preprint]{ChenDinhFanLing}.
	Shortening the last coordinate gives the self-orthogonal
	$[7,3,5]_7$ code $\cC_g(7)$.
	At length nine, $x^8+1$ gives minimum distance two
	by Corollary~\ref{cor:maximal-length}, and
	$\langle g,x^4g\rangle=1$.
\end{example}

\section{Conclusion}\label{sec:conclusion}

We have classified the monic generators $g$ for which
$\cC_g(r+k)$ is Euclidean self-orthogonal and GRS, with
$3\le k\le r$.
For fixed $r,k$, these generators form a finite set over each
algebraically closed field.
The classification determines their maximal self-orthogonal
lengths and the associated shortening relations.
Frobenius stability further gives their least coefficient fields
and exact counts over finite fields.

For $3\le k<r$, the same model contains infinitely many
monomial equivalence classes of self-orthogonal non-GRS MDS
codes over every algebraically closed field.
For every finite field $\F_q$ and every positive integer $B$,
at least $B$ such codes can be realized over a common finite
extension of $\F_q$, while remaining non-GRS and pairwise
monomially inequivalent over $\overline{\F}_q$.

\section*{Statements and Declarations}

\subsection*{Funding}
This work was supported by the National Natural Science Foundation of China (grant numbers 12571003 and 12501006) and the Basic and Applied Basic Research Foundation of Guangdong Province
(grant number 2024A1515010589).

\subsection*{Competing Interests}
The authors declare no competing interests.

\subsection*{Data Availability}
No external datasets were used in this study.

\subsection*{AI Use Declaration}
During the preparation and revision of this manuscript,
generative AI tools were used for literature searches,
comparisons with related work, and finding examples.
The mathematical claims, proofs, and final presentation
remain the responsibility of the authors.

\end{document}